\pdfoutput=1
\documentclass[
    ,final 
  ]
  {aipproc}

\layoutstyle{8x11double}

\begin{document}

\title{Coronal Pseudostreamers: \\
Source of Fast or Slow Solar Wind?
}


\author{Olga Panasenco}{
  address={Helio Research, 5212 Maryland Ave, La Crescenta, CA 91214}
}

\author{Marco Velli}{
  address={Jet Propulsion Laboratory, California Institute of Technology, Pasadena, CA 91109}
}

\begin{abstract}
 We discuss observations of pseudostreamers and their 3D magnetic configuration as reconstructed with potential field source surface (PFSS) models to study their contribution to the solar wind. To understand the outflow from pseudostreamers the 3D expansion factor must be correctly estimated. Pseudostreamers may contain filament channels at their base in which case the open field lines diverge more strongly and the corresponding greater expansion factors lead to slower wind outflow, compared with pseudostreamers in which filament channels are absent. In the neighborhood of pseudostreamers the expansion factor does not increase monotonically with distance from the sun, and doesn't simply depend on the height of the pseudostreamer null point but on the entire magnetic field configuration.
\end{abstract}

\maketitle

\section{INTRODUCTION}

  The correlation between coronal holes and high-speed solar wind was interpreted by Levine, Altschuler and Harvey [1] in terms of expansion of magnetic flux tubes: higher speed winds corresponding to slower expansion. Following this interpretation Wang and Sheeley [2] created an empirical model (WS model) for the solar wind speed based on simulations of the observed daily wind speeds over the interval 1967-1988. They found that regions with multiple coronal holes of the same polarity can became sources of high-speed wind by restricting the volume that the coronal hole flux can occupy, and lowering the associated expansion factor. Such structures, which overlie twin loop arcades and separate holes of the same polarity are known as pseudostreamers [3, 4]. 
  
  Pseudostreamer-like structures show a wide spectrum of scales, their base and null-point height can vary from less than 100 Mm up to 700 Mm [3, 4, 5, 6]. Recent observations and PFSS modeling show that the solar wind from pseudostreamers can be not only fast but also slow and intermediate or 'hybrid' type [7]. Advocates of pseudostreamers as prominent sources only of slow wind reach their conclusion based on observations of what is in reality a small sample from the whole range of pseudostreamers [8, 9]. Wang et al. [7], have done a more careful work on analyses of the wind-pseudostreamer link. They conclude that pseudostreamers produce a 'hybrid' type of outflow that is intermediate between classical slow and fast solar wind.  
   \begin{figure}
  \includegraphics[height=.35\textheight]{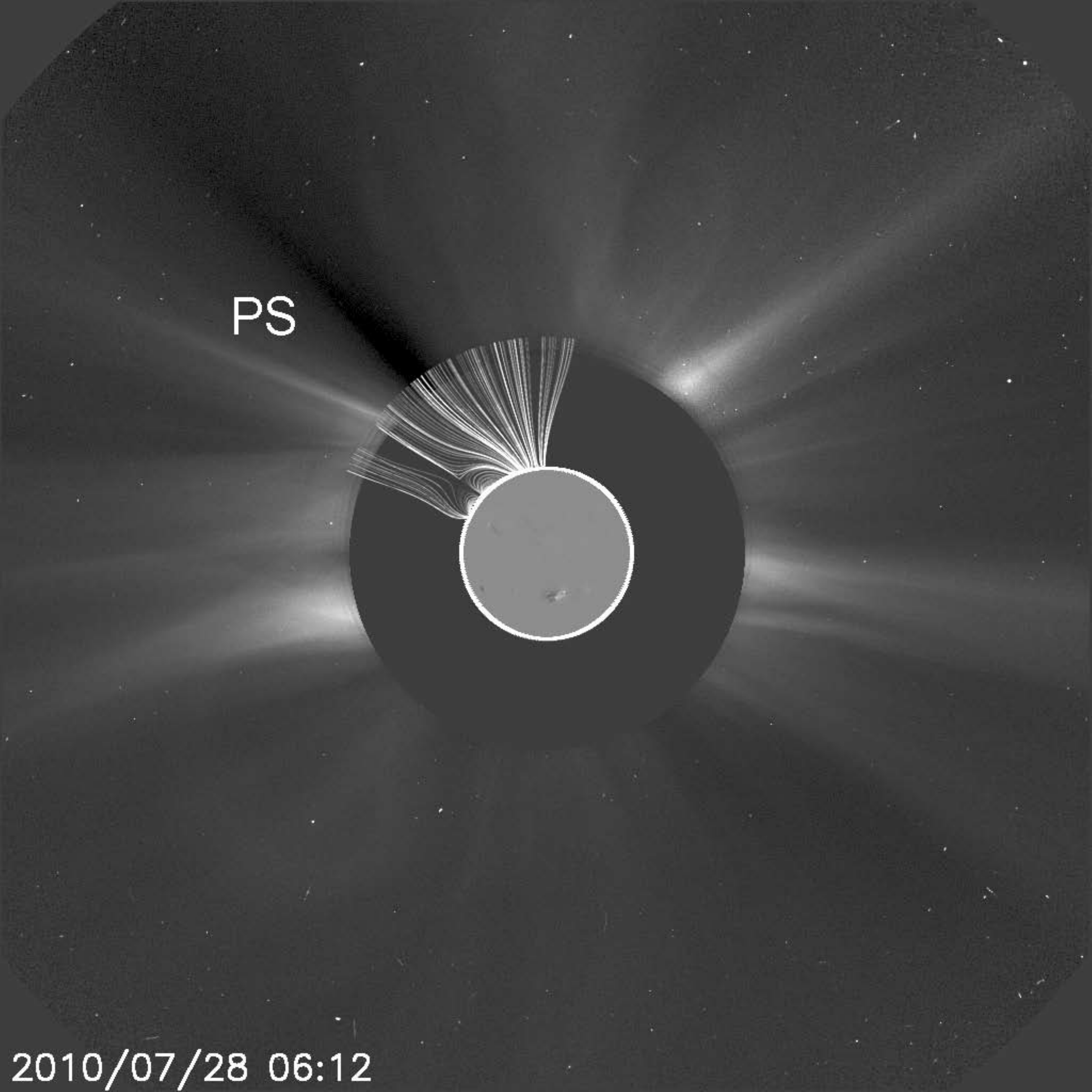}
  \caption{The white light image of a pseudostreamer (PS) at the northeast limb as observed with the SOHO/LASCO C2 coronagraph at 06:12 UT on 2010 July 28. PFSS calculated field lines have been inserted inside the occulting disc to show the pseudostreamer geometry.}
\end{figure}
 \begin{figure}
  \includegraphics[height=.28\textheight]{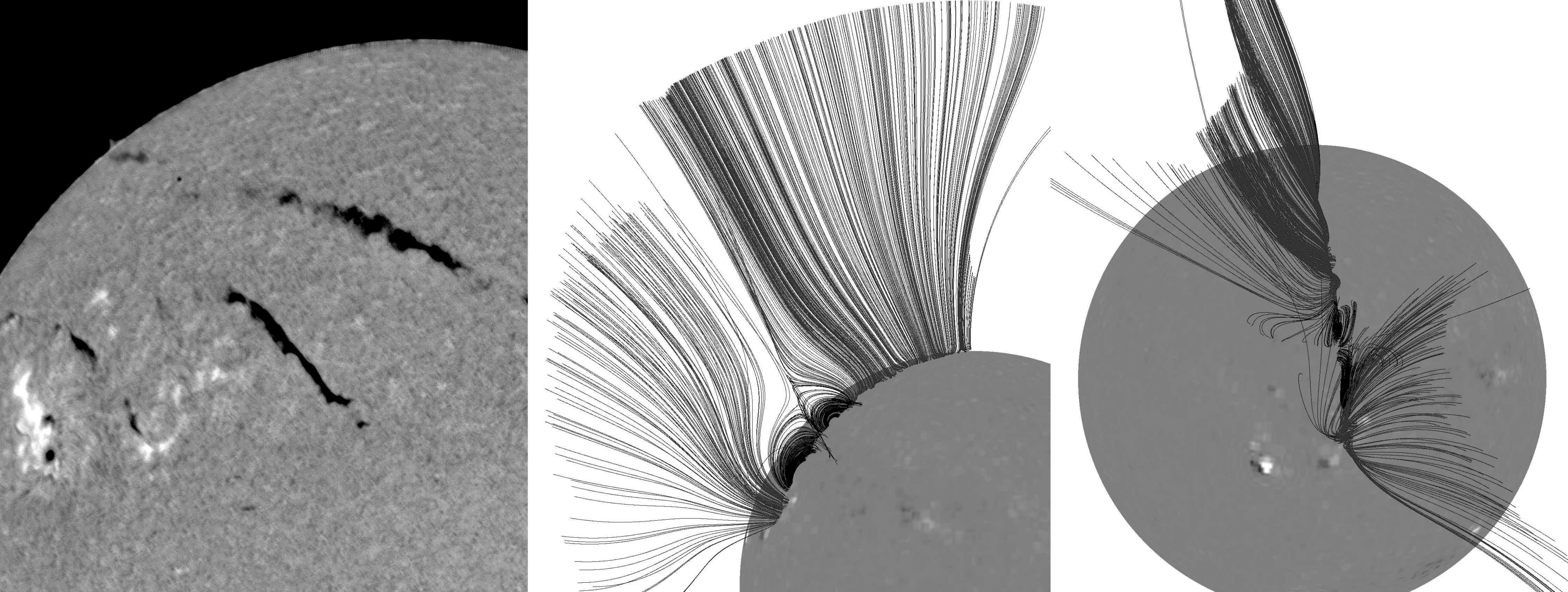}
  \caption{Twin filaments observed in H-alpha (BBSO) on 2010 July 30 at 16:02 UT with the corresponding magnetic field PFSS extrapolation (SDO/HMI) above the top two dextral filaments: limb view (left), top view (right). Note the separation of field lines on opposite sides of the pseudostreamer separatrix-skeleton. The null-point is $\sim$  300 Mm high, the PS base is $\sim$ 450 Mm wide.}
\end{figure}
  However, all of the above studies used the observed 2D projection onto the plane of the sky as an indicator of pseudostreamer expansion, limiting the generality of the their conclusions. The areal expansion of a flux tube is determined, mathematically, from the property of zero divergence of the field. This may be written in terms of a field line integral expressing the conservation of flux for a bundle of field lines as
\begin{equation}
A(P)/ A(0)=exp\bigl(-\int_0^P {ds\over B} {\vec B\over B} \cdot \nabla B \bigr)
\end{equation}
where $\vec B$ is the magnetic field, $B$ its absolute value, and the integral is taken along a field line from $0$ to $P$ with $ds$ being the increment. Its evaluation therefore depends on knowledge of the vector field at each point along a field line, and depends on the way the field expands in space in 3D.

The present paper argues that this true 3D expansion factor, with its dependence on the height of the pseudostreamer dome-fan separator null-point, divergence of pseudostreamer branches in latitude and global magnetic configuration should be taken in account to estimate wind speed. We will discuss observations of pseudostreamers and their 3D magnetic configuration as reconstructed with a potential field source surface (PFSS) model. The latter  [10, 11] uses as a lower boundary condition the magnetic field derived from an evolving surface-flux transport model. This is an extension of the traditional surface-diffusion model [12, 13, 14] and incorporates all of the ingredients that have been demonstrated to play a role thus far: flux emergence, random-walk dispersal, meridional advection, differential rotation, and removal of flux via cancellation. It incorporates the observed MDI and HMI line-of-sight magnetograms but is not a simple synoptic extension of them. This is an important point to stress since the model identifies coronal features, such as the coronal imprint of filament channels, which seems surprising given the non-potentiality of such structures (discussed further below).

\section{CORONAL PSEUDOSTREAMERS AND FILAMENTS}

Coronal pseudostreamers appear in globally unipolar regions above multiple polarity reversal boundaries and can be observed at the solar limb as thin bright rays (Figure 1). Smaller scale pseudostreamer configurations can support jets, or polar plumes [6]. Some of the polarity reversal boundaries below pseudostreamers can be filament channels. When this is the case both polarity inversion lines turn out to be filament channels, often containing what we have named \emph{twin filaments}, of the same chirality [15]. The magnetic structure of pseudostreamers for cases with and without twin filaments, as reconstructed using the PFSS technique, is significantly different. Branches of pseudostreamers on opposite sides of the separatrix surface diverge when filaments are present, in association with the strong horizontal component of the field present in filament channels. 

The presence of strong shear in the confined cavities underneath pseudostreamers containing filament channels is a natural consequence of filament channel formation (e.g., [16]). It is remarkable that this shear component remains as a strong signature of the underlying twin filament channels in our PFSS reconstruction even above and outside the cavities, given the presumed nature of the filament channel as containing strong currents. Yet there must be a difference in the distribution of photospheric flux between a filament channel neutral line and an arbitrary casual one. Indeed, photospheric shear and flux transport models correctly describe filament channel formation statistically based on photospheric transport [17]. A plausible important effect is the finite longitudinal extent of filament channels - so that the polarities on opposite side of the inversion line are not translationally invariant along it. 

Figure 2 shows the PFSS extrapolation of the magnetic field measured by SDO/HMI of the tripolar configuration above two dextral filaments observed 2010 July 30. The strong shear associated with twin filaments leads to separation of field lines on opposite sides of the pseudostreamer separatrix-skeleton leading to a greater expansion factor above the filament channels. The direction of the diverging open field lines around the pseudostreamer corresponds to the direction of the arcade skew above the dextral filaments [18].
\begin{figure}
  \includegraphics[height=.20\textheight]{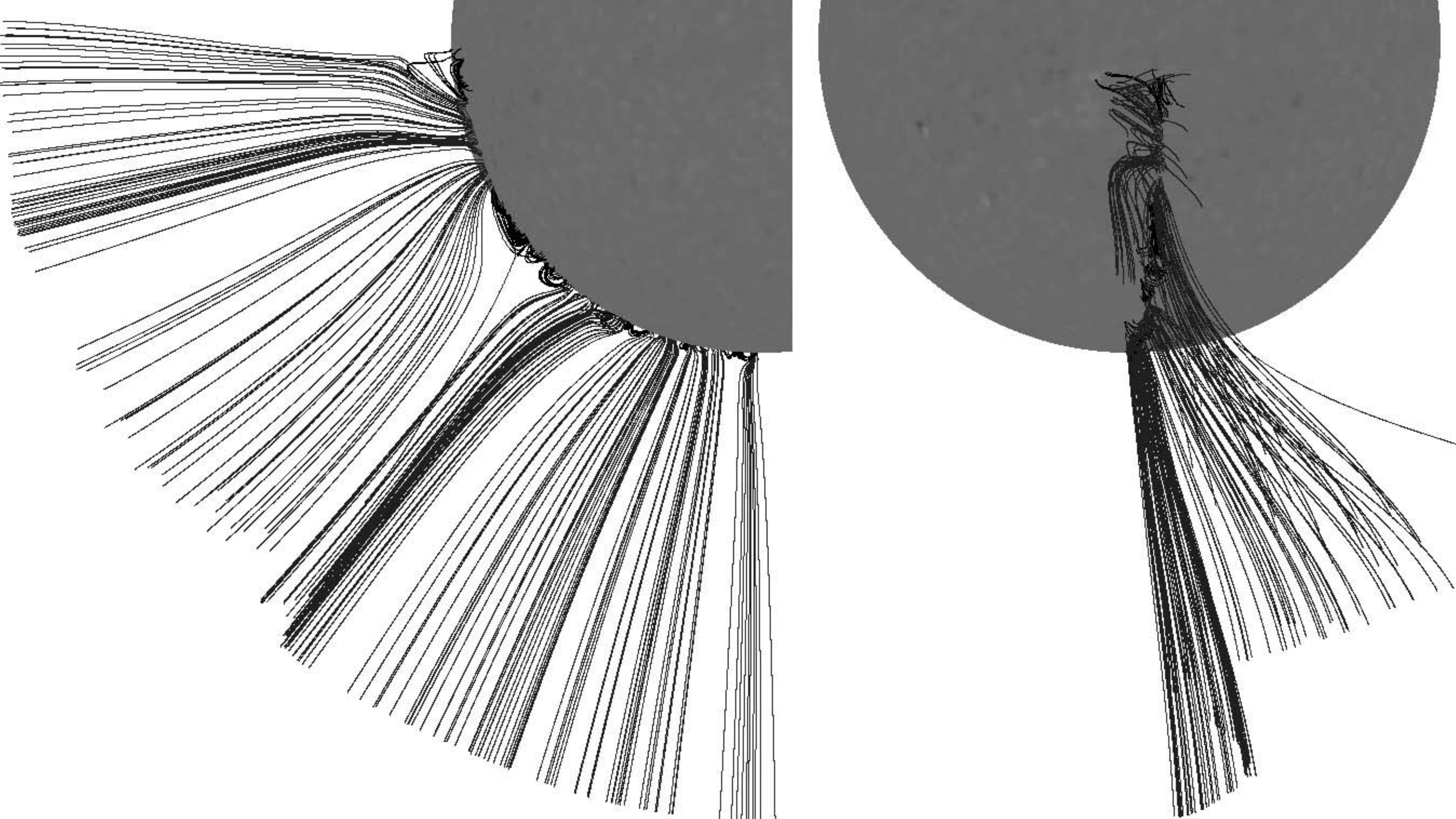}
  \caption{The PFSS model of a pseudostreamer without filament channels in the southern hemisphere. The length of the PS base is $\sim$ 300 Mm, the null-point height is $\sim$ 120 Mm, the PS branches converge, the expansion factor is very low ($\sim$ 7) and the solar wind speed was $\sim$ 600 km/sec (2008 July 17, 18:04 UT, CR 2072).}
\end{figure}
As seen from Figure 2, the expansion factor can appear to be very small in the 2D limb projection view but still turns out to be very high when the reconstructed field lines  are viewed from the top: lines expand widely in latitude in opposite directions. 

In the absence of filament channels below a pseudostreamer however, the branches of the pseudostreamer converge. An example of such a  pseudostreamer with a very low expansion factor, shown in Figure 3,  corresponds to solar wind speed $\sim$  600 km/sec (for the ACE spacecraft measurements, see [8]). Such a reconstruction supports the WS model when the 3D expansion factor is taken in account. 

\section{EXPANSION FACTOR AND  NULL-POINT HEIGHT }

\begin{figure}
  \includegraphics[height=.37\textheight]{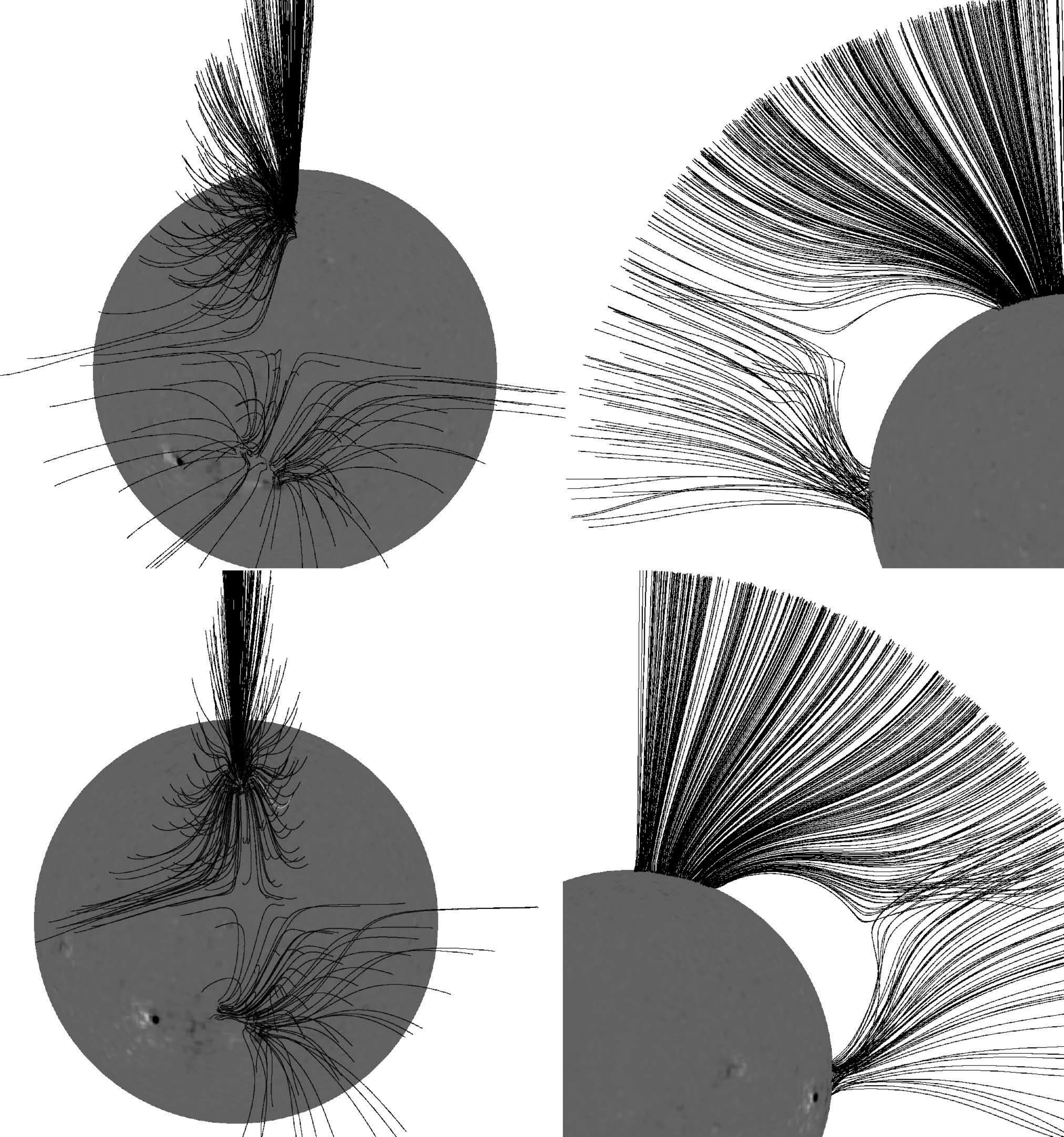}
  \caption{3D PFSS extrapolation of the pseudostreamer magnetic field for CR 2060, 2007 Aug 23 (top row) and Aug 26 (bottom row): branches of the pseudostreamer diverge (left column - top view, right - limb view). The PS base width is $\sim$ 800 Mm, the null-point height is $\sim$ 350 Mm, the expansion factor is > 20.}
\end{figure}
Figure 4 shows the 3D PFSS extrapolation of the pseudostreamer magnetic field observed on Aug 23 and 26, 2007 (CR 2060), during solar minimum. The limb projections (right column) show a very small expansion factor, but the top view (left column) reveals a strong divergence of the pseudostreamer branches, the 3D expansion factor estimated to be > 20, with a null-point height $\sim$  350 Mm; at 1AU the solar wind speed was $\sim$  350-400 km/sec (ACE measurements, see [19]). These measurements are in agreement with WS model, predicting slow solar wind for such expansion factors, even if the origin is the open field lines from the unipolar region. 

The height of the pseudostreamer null point depends on the relative location and intensity of minor and dominant polarity fluxes, including the distance between coronal holes (of the same polarity) and the presence or absence of filament channels at the pseudostreamer base. Near solar minimum filament channel arcades have a tendency to be taller than during more active solar periods due to the generally weaker fields and the slower canceling process, responsible for building up the whole filament channel system, at photospheric levels. A consequence is that the expansion factor will generally be much higher and the solar wind velocity much lower for such periods.

Figure 5 shows an example of a pseudostreamer with a very low null-point height $\sim$  100 Mm during solar maximum. From 2D limb projections one would expect high-speed solar wind from this location since the apparent expansion factor and null-point height are both very low. However, when viewed from the top, the pseudostreamer branches diverge strongly in latitude, creating a fan-like structure which merges into the heliospheric current sheet (HCS) located eastward from the pseudostreamer and parallel to its base stretched in the north-south direction. The expansion is thus quite large ($\sim$15-20, and the solar wind speed was $\sim$  400 km/sec [7].
\begin{figure}
  \includegraphics[height=.25\textheight]{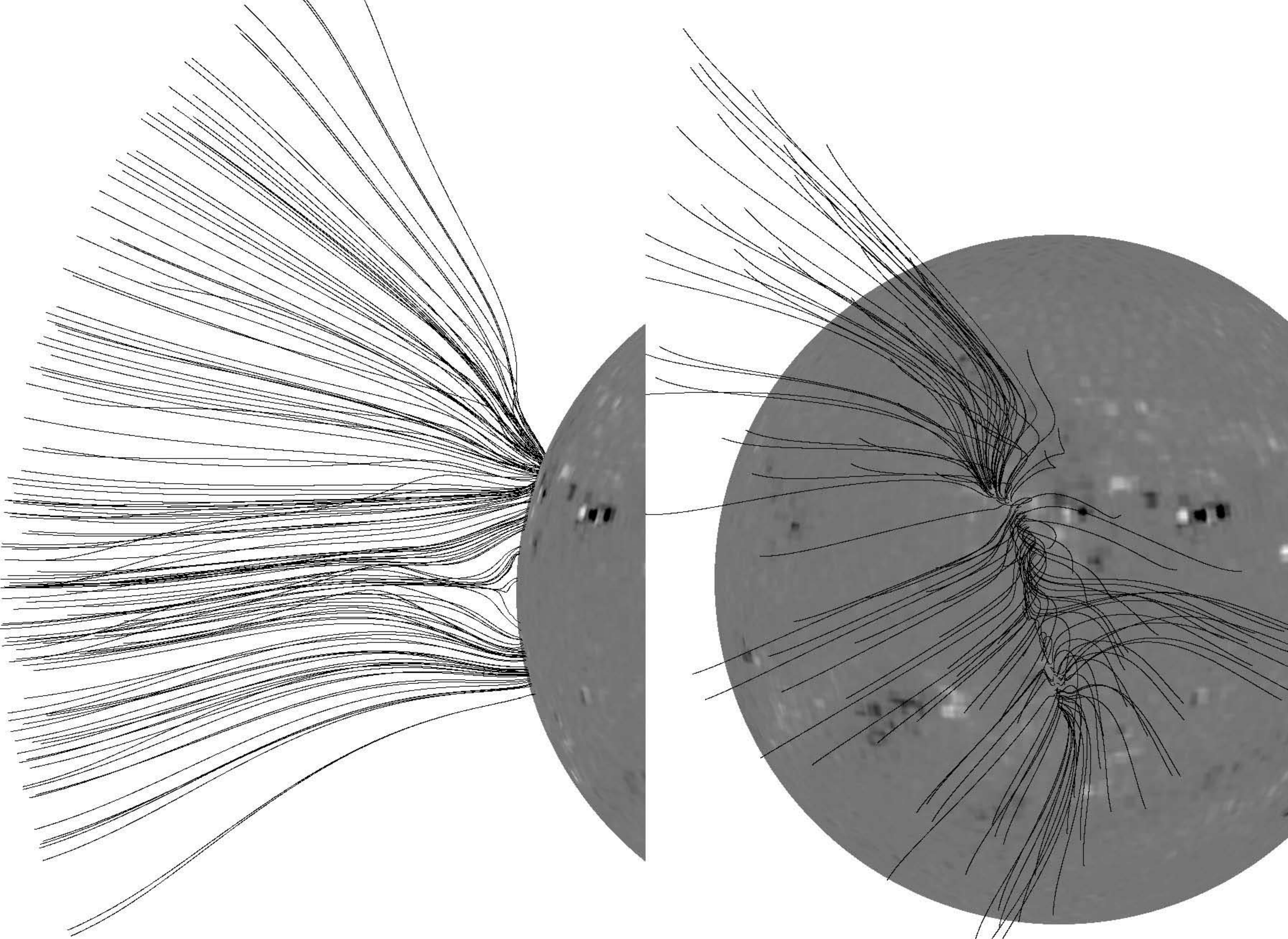}
  \caption{CR 1953. 1999 Sep 05, 18:04 UT. Limb and Earth views of PFSS magnetic field lines at a pseudostreamer near the equator. The null-point height is $\sim$100 Mm, the PS branches diverge, the expansion factor is $\sim$20, the wind speed $\sim$ 400 km/sec.}
\end{figure}

A similar example is shown in Figure 6. Here the heliospheric current sheet plays an important role, leading to an increased expansion factor for the pseudostreamer at greater heights and latitudes. The complex unipolar region, containing three areas with the open magnetic filed, shows changes in the expansion factor with height and latitude. The northern part has very small expansion factors below 100 Mm, but above the field lines diverge in latitude very fast, influenced by the nearly vertical position of the heliospheric current sheet (tilt angle $\sim$  73 degrees). The southern part of this region is squeezed on the eastern side by the same HCS, the expansion factor for most field lines is high ($\sim$ 20) compared to that observed in projection. The solar wind speed was $\sim$  400 km/sec [7].

These examples show that the expansion factor is, in general, very non-monotonic and does not depend directly on the pseudostreamer null-point heights.  The global magnetic configuration and the position of the HCS together with solar activity cycle should all be taken into account.
\begin{figure}
  \includegraphics[height=.26\textheight]{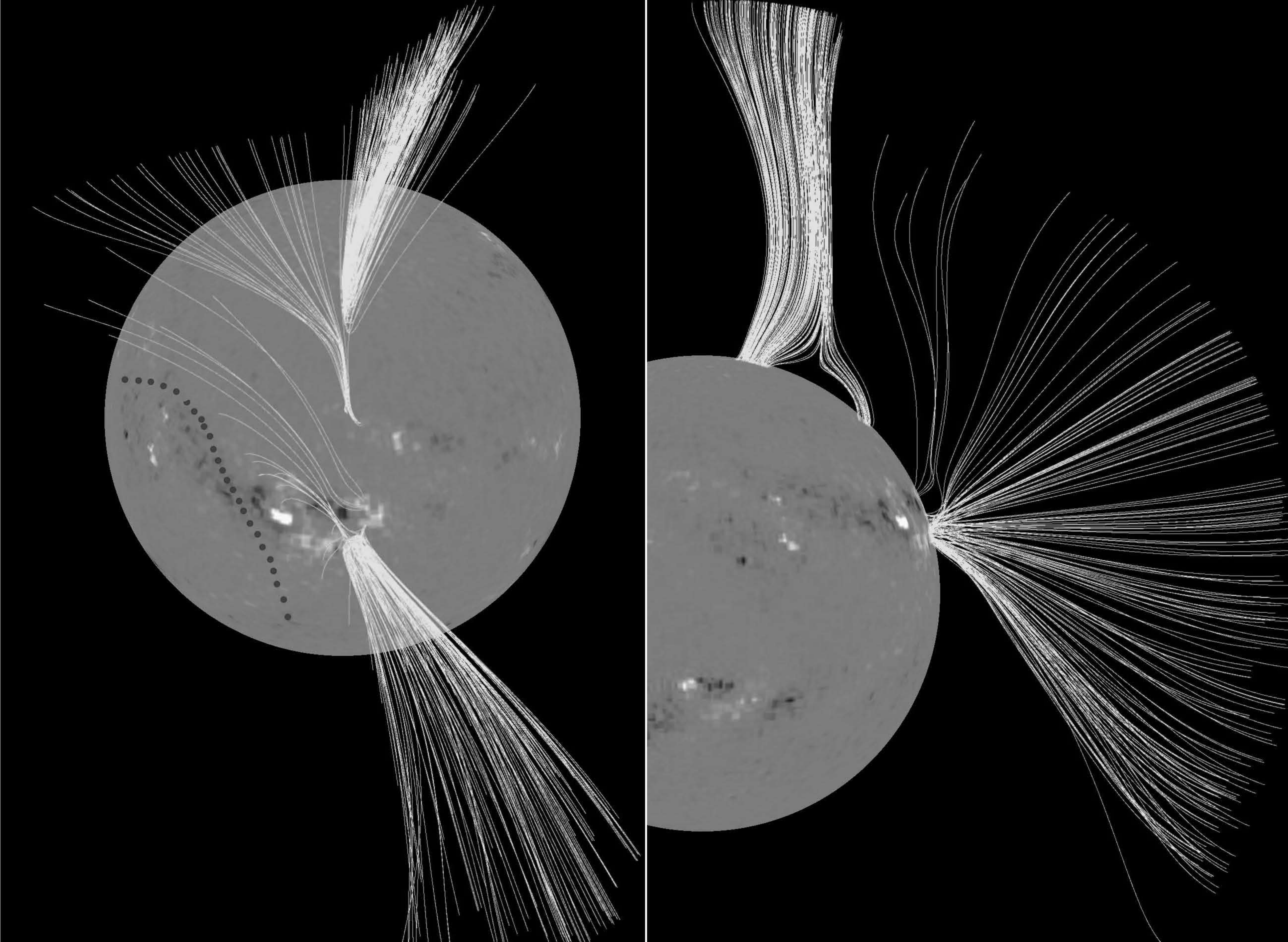}
  \caption{Unipolar open field lines. 1999 Aug 18, 18:04 (PFSS reconstruction: top and limb views). The dark gray dotted line represents the projection of the heliospheric current sheet, the computed tilt angle is $\sim$ 72-73 degrees.}
\end{figure}

\section{DISCUSSION}

Pseudostreamers containing filaments most comonly appear between a polar coronal hole and a small high latitude coronal hole. Such a setup offers at least two neutral lines potentially harboring twin filaments (filaments with the same chirality). The presence of filament channels implies large shear in the whole pseudostreamer system. The branches of the pseudostreamer have a tendency to follow this shear and diverge in 3D. Near solar activity minimum filament channel arcades have a tendency to be taller than for more active solar periods so the expansion factor is going to be much higher for such periods and the solar wind velocity - much lower.

The expansion factor is, in general, very non-monotonic and does not depend only on the pseudostreamer null-point heights.  
Whether a pseudostreamer produces fast or slow wind therefore depends on many factors:
\begin{enumerate}
\item 3D expansion factor (global and local);
\item Presence or absence of filament channels in the pseudostreamer base;
\item Height of the null point;
\item Position of the heliospheric current sheet;
\item Global magnetic configuration on the sun which depends on activity cycle. 
 \end{enumerate}
 
Therefore, general results on the types of wind coming from pseudostreamer type coronal configuration gleaned from  a limited set of observations (such as [9]) should be taken with a grain of salt. As a final note, we remark that well-developed filament channel(s) at  pseudostreamer bases show that efficient interchange reconnection near the pseudostreamer - coronal hole boundaries is not occurring, at least during the lifetime of the filaments (which otherwise could not form, as reconnection to open field lines would remove shear from the configuration).

\begin{theacknowledgments}
  O. P. is  supported in this research under the NASA grant NNX09AG27G. The work of M.V. was conducted at the Jet Propulsion Laboratory, California Institute of Technology under a contract from the National Aeronautics and Space Administration.   
\end{theacknowledgments}

\end{document}